# Training and Turnover in Organizations

Natalie S. Glance, Tad Hogg and Bernardo A. Huberman

Xerox Palo Alto Research Center
3333 Coyote Hill Road
Palo Alto, CA 94304, U.S.A.
glance/hogg/huberman@parc.xerox.com

**Abstract**

We present a two-level model of organizational training and agent production. Managers decide whether or not to train based on both the costs of training compared to the benefits and on their expectations and observations of the number of other firms that also train. Managers also take into account the sum of their employees' contributions and the average tenure length within their organization. Employees decide whether or not to contribute to production based on their expectations as to how other employees will act. Trained workers learn over time and fold their increased productivity into their decision whether or not to contribute. We find that the dynamical behavior at the two levels is closely coupled: the evolution of the industry over time depends not only on the characteristics of training programs, learning curves, and cost-benefit analyses, but on the vagaries of chance as well. For example, in one case, the double dilemma can be resolved for the industry as a whole and productivity then increases steadily over time. In another, the organizational level dilemma may remain unresolved and workers may contribute at fluctuating levels. In this case the overall productivity stays low. We also find a correlation between high productivity and low turnover and show that a small increase in training rates can lead to explosive growth in productivity.

## Introduction

During periods of slow growth and a weak economy, corporations often cut programs in order to maintain profitability. Training programs in particular are often targeted because employee turnover is generally higher during times of economic uncertainty [1]. Even in the best of times, organizations must decide how much to invest in on-the-job training, balancing the benefits of increased productivity against the costs of training. Even worse, since trained workers can migrate more easily between competing firms, another firm can potentially enjoy the increased productivity of workers trained by the former employer without paying the costs. Consequently, a company's incentive to train may fall when employee turnover is high [2, 3]. Ironically, numerous studies have shown that untrained workers change jobs more often [1, 4]. Can the vicious circle be broken?

An organization's decision whether or not to train its workers also affects the economy, whether or not the firm factors this variable into its decision. If all firms within an industry fail to train their workers then the whole economy suffers. In this sense, training workers is another type of public good [1], a category which encompasses social dilemmas ranging from the support of public radio to the so-called "tragedy of the commons" [5] to recycling programs. Resolving the dilemma depends not only on the benefits and costs associated with a particular training program, but also on the firm's expectations concerning employee turnover and the policies of competing firms.

Employees face a similar dilemma in their choice of how much they contribute to the overall productivity of the organization. If employees receive a share of the profits regardless of their contribution, some may decide to free ride on the efforts of their fellow workers. However, if all employees reason similarly, the company will fail. Profit-sharing and employee-ownership exacerbate the dilemma. While, in principle, the problem could be resolved by strict management, in practice, worker monitoring is always imperfect and employee effort can vary from high to low within the range allowed.

The two dilemmas on the employee and organizational levels are tightly interlaced. On the one hand, the benefits of training accrue only to the extent that employees contribute to the organization. Thus, a firm should take into account how it expects a training program to affect employe effort as well as employee turnover. On the other hand, trained workers produce at higher rates, which in turn may affect how much they contribute and how often they migrate to other firms compared to untrained workers. (Once again, all the makings for a vicious circle!)

In this paper we study the dynamics of training and turnover in firms facing both organizational and employee-level dilemmas. First we establish a simple model that captures these conflicts and incorporates imperfect information and both worker and organizational expectations. Organizations can be both created and dissolved, and employees can move between firms, start new ones, or leave the industry for good. Next we summarize the different ways the dilemmas can unfold over time, collated from a number of computer experiments. For example, under one set of conditions, the double dilemma can be resolved for the industry as a whole and productivity then increases steadily over time. Alternatively, the organizational level dilemma may remain unresolved and workers may contribute at fluctuating levels. In this case the overall productivity stays low. We also find a correlation between high productivity and low turnover and show that a small increase in training rates can lead to explosive growth in productivity.

Our dynamical model of training and turnover in organizations both confirms the empirical observation that the two variables are tightly interlinked and reveals how the connections might be unraveled.

## Modelling Organization and Agent Strategies

In this section, we describe our model of organizational training, individual learning, and decision-making on both the individual and organizational levels. In our simplified model, all organizations within an "industry" produce the same good, for which there is a completely elastic demand outside the industry. This assumption means that the industry can grow indefinitely since there is no ceiling for production. Agents, or "employees," can move between organizations, within the bounds allowed by the organizations' "managers." The managers must decide whether or not to train the agents in their own organization, and the agents must decide whether or not to contribute to production.

### Interwoven social dilemmas

Our model of management training and employee production is a two-level social dilemma. At the level of the agent, each individual must decide whether or not to contribute to production (a binary approximation to the continuous range of effort they can deliver). For the case of profit-sharing assumed by the model, each agent receives an equal share of its organization's total production, independent of its contribution. Each agent is tempted to free ride on the industriousness of the other agents, but if all agents do so, nothing is produced and everyone loses.

On the higher level of management, organizations must decide whether or not to train their agents. If a manager decides to train, then members of its organization learn over time, and when its members do contribute to production, they do so at progressively higher levels as time progresses. However, training agents comes with a cost to the total utility produced by the organization, which management must take into account. Why should an organization train its agents only to have them stolen away by a competitor? On the other hand, if all agents receive training, the entire industry is better off, collecting higher and higher utility over time. Thus the dilemma.

### Expectations

Recent work on the dynamics of single organizations suffering from the agent-level social dilemma has shown that high levels of production can be sustained when groups are small or hierarchically structured into smaller groups with fluid boundaries [6–8]. The ongoing nature of the social dilemma lessens its severity if the agents take into account the future when making decisions in the present. How an agent takes into account the future is wrapped into what we call its expectations. The barest notion of expectations comes from the economic concept of horizon length. An agent's horizon length is how far it looks into the future, or how long it expects to continue interacting with the other agents in its organization. The agent's horizon may be limited by its lifetime, by its projection of the organization's lifetime, by bank interest rates, etc.

Here our notion of expectations parts from the standard rational expectations treatment in economics [9]. Rational expectations assumes that agents form expectations about the future using near-perfect knowledge of the underlying model. This notion is self-consistent, but circular: the agents predict the future exactly.

We formulate expectations more as beliefs than knowledge. (In a more complex model the beliefs underlying the expectations could evolve over time as more information became available.) In our model of expectations, agents believe that their present actions will affect those of others in the future. The extent of the effect depends on the size of the organization and the present level of production. The larger the group, the less significance an agent accords its actions: the benefit produced by the agent is diluted by the size of the group when it is shared among all agents. If an agent free rides, it can expect the effect to be very noticeable in a small group, but less so in a larger group. This is similar to the reasoning a student uses when deciding whether or not to attend a lecture she would prefer to skip. Among an audience of 500, her absence would probably go unnoticed (and if all students in the class reason similarly . . . ). On the other hand, in a small seminar of ten, she might fear the personal censure of her professor.

In our model, the agents also expect that their actions will be imitated by other agents and that the extent of this mimicry depends on present levels of production. An agent expects that if it decides to free ride ("defect") in a group of contributors, or "cooperators," others will eventually choose to defect as well. The agent also believes that the rate at which the switchover occurs over time depends on the fraction of the group presently cooperating. The more agents already cooperating, the faster the transition to defection. Similarly, an agent expects that if it starts cooperating in a group of free riders, others will start cooperating over time. Once again the agent believes that the rate depends on the proportion of cooperators, which in this case is very low. Our key assumption is that agents believe their actions influence contributors, or "cooperators" more than sluggards, or "defectors."

Consider the set of beliefs the agent expects of others in the context of recycling programs. Not too long ago very few towns had such programs. Perhaps you would read in the paper that a small town in Oregon had started a recycling program. Big deal. But several years later, when you read that cities all over your state have jumped onto the recycling bandwagon, then suddenly the long-term benefits of recycling seem more visible: recycled products proliferate in the stores, companies turn green, etc. Alternatively, imagine some futuristic time when everyone recycles, in fact your town has been recycling for years, everything from cans to newspapers to plastic milk jugs. Then you hear that some places are cutting back their recycling efforts because of the expense and because they now believe that the programs don't do that much good after all. You think about all your wasted effort and imagine that the other towns still recycling are reaching the same conclusion. Suddenly, your commitment to recycling no longer seems so rational.

To sum up, agents believe that the strength of their influence on the amount of cooperation decreases with the size of the group, increases with the current proportion contributing, and extends into the future as far as their horizon. To some extent, this set of beliefs is arbitrary and, certainly, domain-specific. It is easy to imagine other scenarios for which another set of expectations would be more appropriate. However, there is a class of expectations for which the general conclusions of our work hold.

In the present interlocking model of organizational training and agent cooperation, we extend the formulation of expectations to the organizational level. Managers decide to train or not based on the number of organizations in the industry and on the number that presently train their agents. Folded into this decision and into their expectations is the behavior of the agents that comprise a manager's organization. A manager's horizon length depends on the tenure lengths of its agents: the longer its agents stick around, the longer it expects them to stick around in the future, and the more reason a manager has to train them. Likewise, a manager predicts greater future value from training when more of its agents are actively contributing instead of free riding.

## Strategies for cooperation

A low-level description of the model includes a number of parameters that describe agent and organizational attributes. The tables below list these parameters and their definitions for agents and organizations.

In a profit-sharing organization in which individual agents receive equal shares of the utility produced by the group, the utility to agent *i* in organization *m* is its share

***Agent attributes***

| | |
|---|---|
| $b_{min}$ | Baseline benefit (per unit time) of cooperation |
| $b_i^m$ | Benefit (per unit time) of cooperation for agent *i* belonging to organization *m* |
| $c$ | Cost (per unit time) of cooperation |
| $H$ | Horizon length |
| $k_i$ | Binary variable: $k_i = 1$ if agent *i* contributes, 0 otherwise |
| $\gamma$ | Learning rate |
| $r$ | Fraction of learning transferred across organizations |
| $t_i^m$ | Tenure length of agent *i* in organization *m* |
| $\alpha$ | Reevaluation rate |
| $p$ | Measure of uncertainty |

***Organizational attributes***

| | |
|---|---|
| $N$ | Total number of organizations in the industry |
| $n_m$ | Number of agents in organization *m* |
| $\kappa_m$ | Binary variable: $\kappa_m = 1$ if organization *m* trains, 0 otherwise |
| $T$ | Training cost per agent per unit time |
| $H_m$ | Horizon length for manager *m* |
| $\alpha_m$ | Reevaluation rate for manager *m* |
| $q$ | Measure of uncertainty |
| $f_c^m$ | Estimated fraction cooperating in organization *m* |

minus its cost for cooperation:

$$U_i = \frac{1}{n_m} \sum_{j=1}^{n_m} b_j^m k_j - c k_i. \tag{1}$$

The individual utility depends indirectly on the managerial policies of organization *m*. If manager *m* does not train, then the benefit of cooperation for its agents stays fixed, while if the manager does train, the benefit increases linearly over time. The linear model of learning is described by the differential equation

$$\frac{db_i^m}{dt} = \gamma \kappa_m. \tag{2}$$

All agents start off with $b_i^m = b_{min}$. If an agent switches to a different organization *l* in the industry it retains only a fraction of the training it received:

$$b_i^l = r(b_i^m - b_{min}) + b_{min}, \tag{3}$$

but its benefit never falls below the baseline benefit, $b_{min}$. This models incomplete transfer of knowledge between organizations.

The utility to the organization as a whole is the total amount produced by its constituent agents minus any training costs:

$$U^m = \sum_{j=1}^{n_m} b_j^m k_j - n_m T \kappa_m. \tag{4}$$

Notice that the organizational utility per agent has a functional form that is very similar to the individual agent utility. Only the cost term is different.

Agents and managers use their respective utility functions to guide their decisions to contribute or not contribute, to train or not train. They project future earnings in accordance with their expectations and their horizon lengths. For individual agents, the criteria for cooperation was derived in [7] for a simpler model and extends easily to the present case. Individuals cooperate if their observed share of production

$$\langle b \rangle^m = \frac{1}{n_m} \sum_{j=1}^{n_m} b_j^m k_j \tag{5}$$

exceeds the critical amount

$$b_{crit}^m \equiv \frac{b_{min}}{H\alpha} \left( \frac{n_m c - b_i^m}{b_i^m + \gamma \kappa_m H - c} \right). \tag{6}$$

According to this criterion, beyond a critical group size, no agent will cooperate, and below a second critical group size, all agents will cooperate. Between the two limits, there are two equilibrium points, one of mostly cooperation, the other of mostly defection. The group dynamics tends towards the equilibrium closest to its initial starting point. Generally, one of the equilibria is metastable, while the other is the long-term equilibrium. If a group falls into a metastable state, it may remain there for very long times (exponential in the size of the group). Because of uncertainty (modeled using the parameters *p* and *q*) the group will eventually switch over to the global equilibrium very suddenly (in time logarithmic in the size of the group), as shown in [6].

The training criterion for organizations follows by analogy. A manager trains when the observed fraction of organizations training exceeds the critical amount

$$f_{crit}^m \equiv \frac{1}{H_m \alpha_m} \left( \frac{Nt - \gamma f_c^m}{\gamma f_c^m - t} \right). \tag{7}$$

This criterion has the following properties. Managers are more likely to train when their horizon lengths are long, when training costs are low compared to the agents' learning rate, when the number of organizations is small, and when they estimate a large proportion of their agents to be cooperating. A manager can estimate the fraction cooperating from the production level it observes by inverting the organizational utility given by Eq. 4. This estimate will differ from the actual fraction cooperating since an organization's agents may have received different amounts of training and will consequently have different benefits for cooperation. We model the manager's estimate of the fraction cooperating as

$$f_c^m = \frac{1}{b_{min}} \left( \frac{1}{n_m} \sum_{j=1}^{n_m} b_i^m k_i \right). \tag{8}$$

This estimate overstates the amount of cooperation and worsens as the agents learn over time.

We intend the two conditions for action to be taken as heuristic guidelines as opposed to formulae writ in stone. While the agent-level condition for cooperation was derived from the expectations sketched out earlier, its qualitative features are what interest us. We expect the heuristic form of the criteria to hold for a wide range of expectations. For some sets of expectations they may not hold, in which case a different model would then be appropriate.

## Fluidity

We also model the changing structural nature of industries over time. We use the term fluidity to describe the ease with which structure can change. The parameters which govern the amount of fluidity in an industry are listed in the table below. For the purposes of our model, we consider them as given exogenously; they could also be thought of us under the control of some metalevel agent (say, some regulatory mechanism) which adjusts the fluidity parameters in order to optimize the overall utility of the industry or perhaps even under individual agent control.

Fluidity describes the ease with which agents can move within an organization from subgroup to subgroup, how promptly they leave the organization for another one or leave the industry completely seeking higher personal utility, and how readily they start an organization of their own. Organizations restrict structural fluidity to the extent they make it difficult for agents to join and difficult for them to leave or move within their organizations.

In this work, structural fluidity is modeled in the following manner. Managers control the rate at which constituent agents choose to move between organizations and the rate at which agents from a pool of agents exterior to the industry can join, but do not restrict agents from leaving. Specifically, agents move between organizations

***Fluidity parameters***

| | |
|---|---|
| $\mu$ | moving threshold |
| $\eta$ | break away threshold |
| $\Omega$ | entrepreneurial rate |
| $\rho$ | joining threshold |

or join an organization only when invited by a manager. Agents accept or decline the invitation according to moving and joining strategies that optimize utility and take into account moving and joining costs (set at the metalevel). Say that agent *i* in organization *m* is invited to join organization *l*. Agent *i* compares its organization's production level with that of organization *l*. Agent *i* will move only if

$$\langle b \rangle_l - \langle b \rangle_m > \mu b_{min}, \tag{9}$$

where $\langle b \rangle_m$ was defined in Eq. 5 and $\mu < 1$. Similarly, if an agent *j* is invited to join organization *m* from the outside pool of agents, the agent will join organization *m* only if the organization's production level exceeds the agent's costs:

$$\langle b \rangle_m > \rho c, \tag{10}$$

with $\rho > 1$ generally.

Agents can also decide to "break away" or leave the industry for good. In our model, an agent will break away when its organization's production level falls below a lower threshold parametrized by the break away variable $\eta$:

$$\langle b \rangle_m < \eta c. \tag{11}$$

Some (small) fraction of the time, parametrized by the entrepreneurial rate, $\Omega$, the agent will start a new organization within the industry instead of leaving. In this fashion, the number of organizations in the industry can grow over time. The number of organizations also decreases whenever all agents from one organization have left.

In previous work, we described how structural fluidity within a single organization enables agent-level cooperation [7]. In this paper, we assume that the time-scale of structural change on the organizational level is much shorter than on the industry level so that we can ignore intra-organizational fluidity and better pinpoint the effects of training and inter-organizational fluidity.

## Computer Experiments

The simulation of our model runs on two levels: the agent level and the organizational level. Agents wake up asynchronously according to a Poisson process with mean $1/\alpha$. When they wake up, they either (1) reevaluate their decision to cooperate or not according to the condition for cooperation given in Eq. 6; or (2) reevaluate their choice to stay in their organization, or start a new organization, or break away from the industry completely, following the decision function given in Eq. 11.

Each manager *m* also wakes up asynchronously, but according to a Poisson process whose mean, $n_m/\alpha$, depends on the size, $n_m$, of its organization. This reflects both the more ponderous decision-making of larger organizations and the longer time-scales over which organizations reevaluate their decisions compared to agents. When a manager wakes up, it either (1) reevaluates its decision whether or not to train its agents, following the condition in Eq. 7; or (2) invites an agent from a competitor organization to join. In the second case, if the invited agent refuses to join, then the manager will invite an agent from the outside pool to join. The moving and joining conditions for the agents are given in Eqs. 9 and 10. Organizations, of course, prefer to steal away agents from competitors since they most likely produce at higher levels, thanks to training, but agents will switch only if they perceive a gain in personal utility.

This is only one of many ways to simulate such a model. Our experience running similar types of simulations indicates that the most important feature is that the agent and managerial states be updated asynchronously [10], not synchronously, in order to accurately model continuous time. Other features of the algorithm described above have little effect on the qualitative results described in the next section.

## Results

The dynamics on the organizational level mirrors the agent-level description given earlier: when the number of organizations in the industry exceeds a critical number, none train, and when it falls below another critical number, all train. Again, between these two critical sizes there is a middle region in which there are two equilibria: one in which all managers train, and one in which none train. The transition from the metastable state to the global equilibrium may not happen for a time exponential in the number of organizations and is very sudden when it finally occurs. The critical numbers depend on the learning rate of the agents and the training cost for the organizations.

However, for fluid industries in which agents can move in and out of various organizations, the critical regions for cooperation and defection shift for both agents and organizations. For agents, the critical regions shift because the size of their parent organizations changes over time. A small cooperating organization will tend to grow over time because outside agents see its high productivity. If the organization becomes too large and its agents do not receive training, then eventually a transition to overall defection will take place. Once all the agents in the organization are defecting, the group's size will shrink because many (or all) will break away from the industry or move to another organization. At some point, the group will again be small enough to support cooperation. This cycle of cooperation-growth to defection-attrition and back again repeats over and over for each organization when managers do not train. The amount of cooperation within different organizations and their sizes are coupled because of the agents moving between organizations.

The critical regions also shift in time for each organization, depending on how many of its agents cooperate and how long the agents stay in the same organization. Over time, what was originally an unresolvable dilemma for the managers (so none train) becomes resolvable, and eventually the dilemma can disappear completely. The behavioral regions shift (1) as the agents' tenure lengths change; and (2) as the agents' production levels increase. The agents' tenure in a particular organization increases when agents remain loyal to the their parent organization. Generally, agents are loyal when their colleagues cooperate. Tenure lengths are short when few cooperate within an organization since agents will move often or break away. Agents' production levels increase when their parent organizations train them and when the agents cooperate among themselves.

## Dynamics of industry growth

The dynamics of agent and organizational behavior is closely coupled. Cooperation at one level encourages cooperation on the other level and similarly with defection. At both levels, there can be metastable states which trap the industry in lower-performing states (or higher performing states). For certain parameters the industry is in the two-equilibria region on both the agent level and the organizational level. The rest of this paper will concentrate primarily on the behavior of the industry for this regime.

The dynamics of the industry is highly path-dependent. For the same initial conditions and parameter choices, the industry can evolve to a number of different states. Figs. 1 and 2 show a series of snapshots taken from the time evolution of two industries starting from the same initial conditions. Initially, both industries consist of four organizations, with eight agents each. The total number of agents in the industry is printed at the top of the schematic tree. The agents cooperate initially, as indicated by the filled lower-level circles. None of the managers are training, as indicated by the open upper-level circles (the same code, filled circles for cooperation/training and open circles for defection/no training are used for both agents and managers). Both industries grow in size at first since their agents cooperate and new agents from outside the industry are attracted by the high levels of production (increasing the size of the industry as a whole). Once an organization grows too large, its agents switchover to defection and move to another organization or break away completely (decreasing the size of the industry).

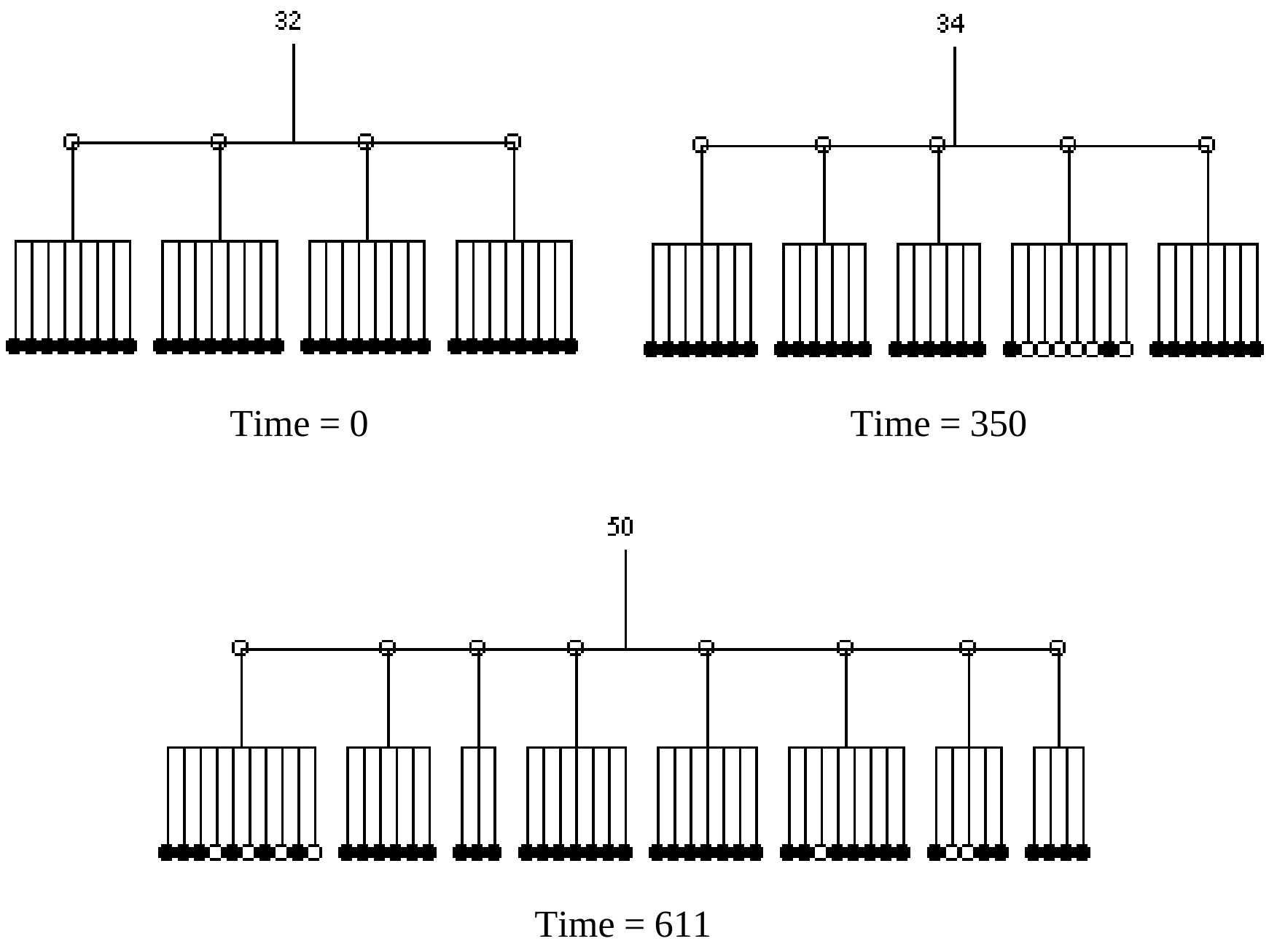

Fig. 1. Snapshots of the time evolution of an industry faced with social dilemmas at both the individual agent and organizational levels. Agents must decide whether or not to cooperate knowing that they receive a share of their organization's production regardless. Organizations must decide whether or not to train knowing that the costs of training will be lost if their agents switch to another organization. The dynamics of the industry is highly path-dependent. For the same initial conditions and parameters, the industry can evolve to a number of different states. The snapshots above are taken from a simulation for which the number of organizations proliferates over time and the dilemma on the organizational level becomes untenable—there is no training of the agents. Since there is no training, the industry's utility can increase only because agents join.

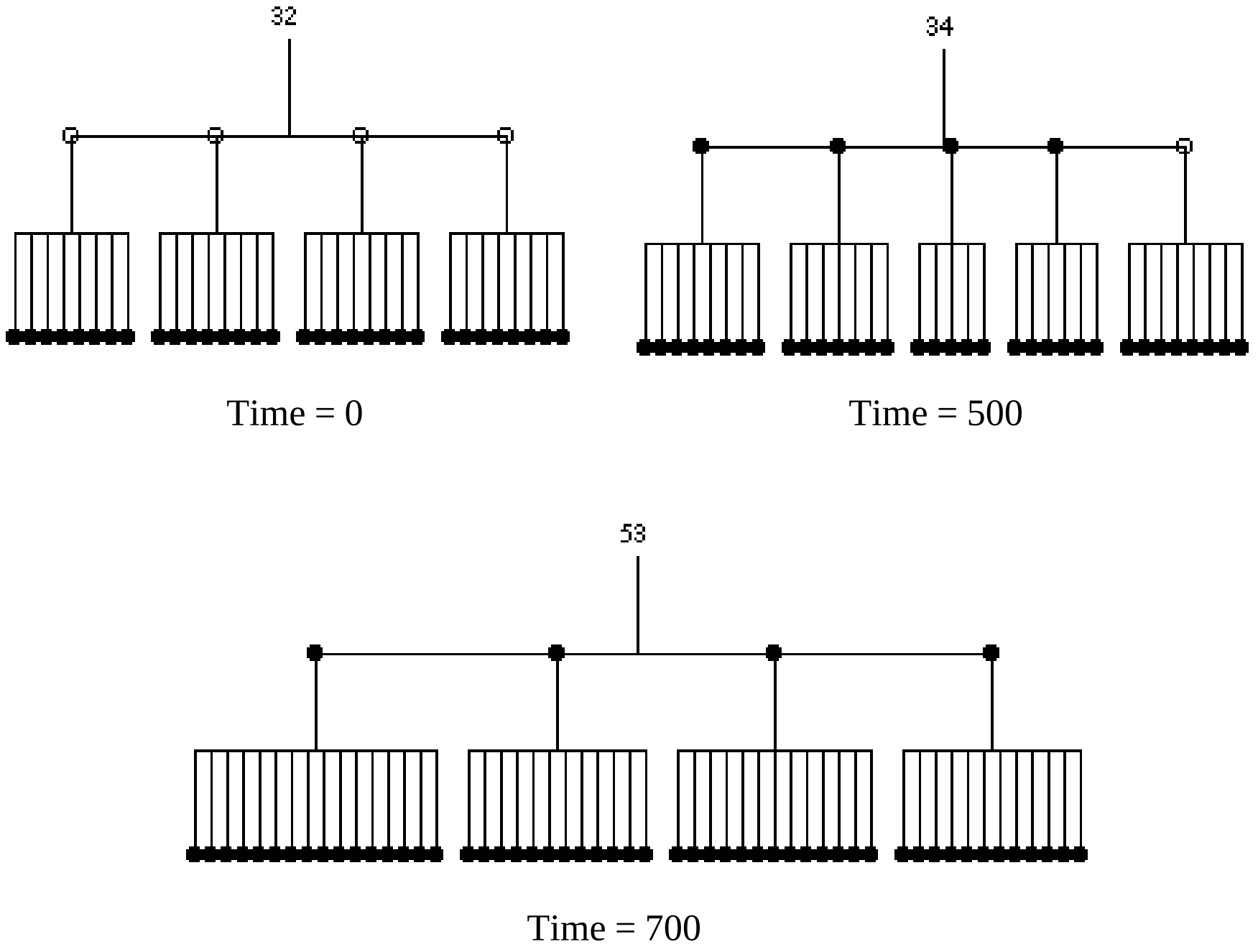

Fig. 2. Snapshots of the time evolution of an industry starting from the same initial conditions and for the same choice of parameters as in the previous figure. The dynamical path followed in this case is very different. The number of organizations remains small for long enough that the organizations switch over to the equilibrium in which all organizations train. Once settled in the training equilibrium, the agents produce at higher and higher levels attracting more agents from outside the organization to join, further increasing the total utility produced by the industry as a whole.

The number of organizations varies stochastically: organizations die whenever all of its constituent members leave, and new organizations form because entrepreneurs strike out on their own. The balance between these two trends depends on the average rates of the various events and on chance. When the number of organizations happens to grow over time, the dilemma on the organizational level becomes untenable—the switchover to overall training never happens. Instead, the number of organizations proliferates over time and the industry tends towards a state of many organizations, each with a small number of members who cycle between states of cooperation and defection. This is the process indicated by Fig. 1. On the other hand, if the number of organizations happens to stay constant or shrink, the managers will eventually all decide to train their agents. In this case, the industry tends towards a state with a small number of very large, highly productive organizations. Fig. 2 depicts such an industry.

The overall utility to the industry over time depends strongly on the path the industry follows. Fig. 3 shows the abrupt deviation in overall utility between the two industries of Figs. 1 and 2. Once the organizations in the second industry switch over to the training equilibrium, the industry's utility rises steadily as the industry attracts more agents who learn and produce more over time.

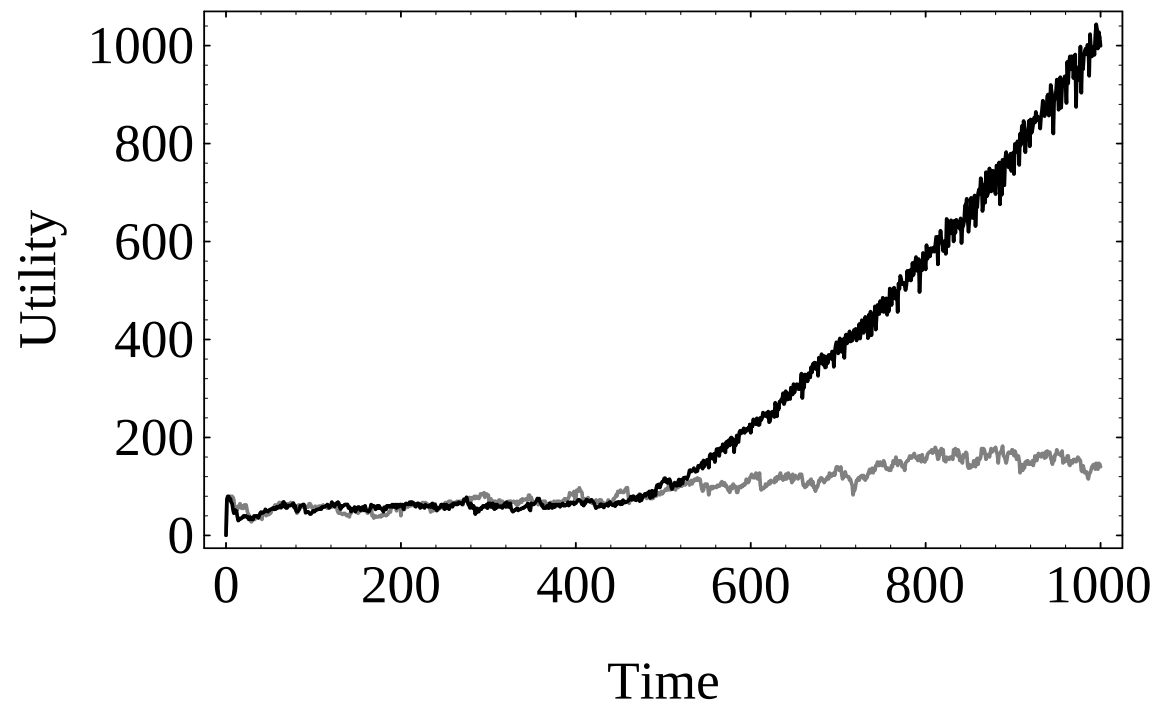

Fig. 3. Utility as a function of time for the two industries described in Figs. 1 and 2, in gray and black respectively. The utility at time step 1000 for the industry of Fig. 1 is more than seven times greater than that of Fig. 2 (1000 compared with 140).

## Maximizing industry-wide productivity

We also studied how sensitive the performance of the industries is to the values of various parameters in the model. We found that two parameters were most significant, given the constraint that the model be kept in the regime of the two-level social dilemma. These were the entrepreneurial rate (the rate at which agents that break away start a new company) and the ratio of the learning rate to the training costs. When the entrepreneurial rate is high, the number of organizations proliferate rapidly and the likelihood that the organizations spontaneously decide to train drops. On the other hand, if the entrepreneurial rate is low, the num-ber of organizations remains small, and the transition to overall training becomes much more likely. Low entrepreneurial rates also limit the overall size of the industry.

The effect of varying the learning rate is more interesting since companies or industries may have some control over this variable. In order to determine the average effect of increasing the learning rate while keeping training costs fixed, we ran the simulation many times for the same choice of parameters and initial conditions. Fig. 4 shows the average utility over 30 runs for each data point. The average utility increases exponentially with increasing learning rates. Increasing the learning rate by less than 50% results in a factor of six explosion in average utility for this set of simulations. Of course, the large increase in utility is the expected value; the actual change in utility for a given industry can vary widely because of the path-dependency described earlier.

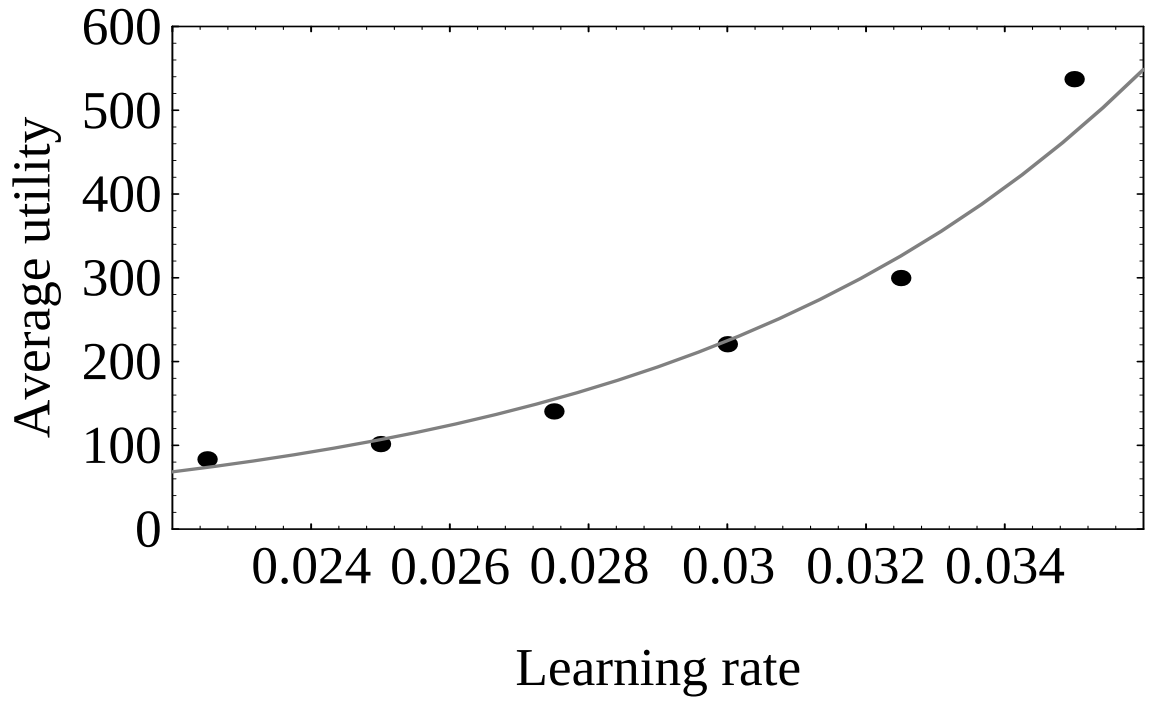

Fig. 4. The average utility produced by an industry over time typically increases exponentially with increasing agent learning rates. The data points were obtained by averaging over 30 runs for each value of the learning rate. The gray curve is an exponential fit to the data.

## The utility of training

Is there a relation between the utility produced by the industry as a whole and the average tenure lengths of its members? This is a very relevant question in today's world of downsizing and rapid turnover. We ran a hundred simulations of the model using the same parameters and initial conditions in order to address this question. Fig. 5 indicates the correlation found between short tenure lengths and lower overall utility for the industry.

## Discussion

In order to understand the interplay of social dilemmas at both the organizational and agent level, we have constructed a simple model that encompasses cost-benefit analyses and expectations at both levels. At the organizational level, managers decide whether or not to train based on both the costs of training compared to the benefits and on their expectations and observations of the number of other firms that also train. Managers also take into account the sum of their employees' contributions and the average tenure length within their organization. At the agent level, employees decide whether or not to contribute to company production based on their expectations as to how other employees will act. When trained, agents learn over time and fold their increased productivity into their decision whether or not to contribute as well.

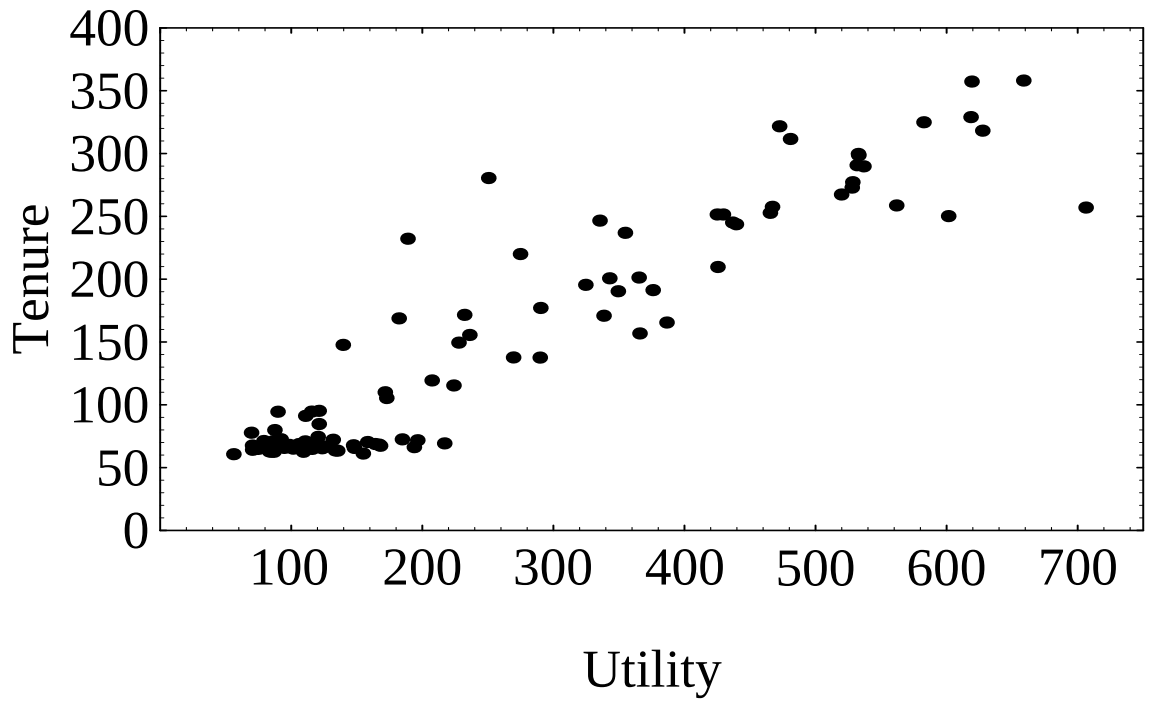

Fig. 5. Scatter plot of average utility versus average agent tenure lengths for 100 simulations of an industry starting from the same initial conditions and identical parameter choices.

We also modeled how easily employees can move between firms, a property we call structural fluidity. In addition, agents can leave the industry for good, and new ones can join. New firms may be created when an agent leaves its parent organization to start a new one. We described how fluidity relieves the dilemma at the agent level by allowing a large, low-productivity organization to break up into smaller pieces. In extreme cases, the organization may dissolve completely. However, when firms break apart in this way, the total number of organizations in the industry proliferates, exacerbating the dilemma on the organizational level.

The dynamical behavior at the two levels is closely coupled because of these interlinked effects. As a result, the dynamical unfolding of the dilemmas on the employee and organizational levels is path-dependent. The evolution of the industry over time depends not only on the characteristics of training programs, learning curves, and cost-benefit analyses, but on the vagaries of chance as well. Starting from the same conditions, an industry can evolve to one of many states. In some cases, it evolves to a stable collection of firms that train their agents and become more productive over time. In other cases, the number of firms proliferates over time, and each firm experiences high worker turnover and low productivity because of the lower contributions of unpaid and, at times, unmotivated, workers.

Our computer experiments also showed a correlation between high turnover and low overall utility to the industry. However, these results were obtained assuming a fixed environment. In the more general case, the environment will change over time, perhaps setting the employees back in their training programs, or bankrupting firms en-tirely. We plan to extend our model to include the effects of a changing environment and to study how the dynamics of the industry will be affected. We expect that in an environment that changes continuously, effectively offsetting some of the benefits of training, that the dilemma on the organizational level will be much harder to resolve. In such a case, organizations that train will not be much better off than those that do, but will still incur the additional costs of training. The effect of an environment that changes intermittently in an abrupt fashion is harder to predict. While the benefits of training will not be as great, we expect there to be a trade-off between training and adaptability.